\documentclass[12pt]{article}
\usepackage{a4wide}
\usepackage{latexsym}
\usepackage{amsmath}
\usepackage{amsfonts}
\usepackage{amscd}
\usepackage{cite}
\usepackage{placeins}

\usepackage{pslatex}
\usepackage{graphicx}
\usepackage[latin1,utf8]{inputenc}
\usepackage[T1]{fontenc}
\usepackage{hyperref}
\allowdisplaybreaks

\newcommand{\bq}{\begin{eqnarray}}
\newcommand{\eq}{\end{eqnarray}}
\newcommand{\eps}{\varepsilon}

\newcommand{\Nletter}{N_{\mathrm{letter}}}

\begin{document}

\thispagestyle{empty}

\begin{flushright}
  TUM-HEP-1521/24 \\
  MITP/24-068 
\end{flushright}

\vspace{1.5cm}

\begin{center}
  {\Large\bf 
The H-graph with unequal masses in quantum field theory \\
  }
  \vspace{1cm}
  {\large Philipp Alexander Kreer${}^{a}$ and Stefan Weinzierl${}^{b}$ \\
  \vspace{1cm}
      {\small \em ${}^{a}$ Physik Department,} 
      {\small \em TUM School of Natural Sciences,} \\
      {\small \em Technische Universit\"at M\"unchen,} 
      {\small \em D - 85748 Garching, Germany} \\
  \vspace{2mm}
      {\small \em ${}^{b}$ PRISMA Cluster of Excellence,} 
      {\small \em Institut f{\"u}r Physik, Staudinger Weg 7,} \\
      {\small \em Johannes Gutenberg-Universit{\"a}t Mainz,}
      {\small \em D - 55099 Mainz, Germany}
  } 
\end{center}

\vspace{2cm}

\begin{abstract}\noindent
  {
We compute the family of Feynman integrals related to the H-graph with unequal masses in relativistic quantum field theory.
We present an $\eps$-factorised differential equation for the 40 master integrals.
The alphabet consists of 29 dlog-forms with algebraic arguments, involving six square roots.
As these square roots are not simultaneously rationalizable,
we express the master integrals in terms of iterated integrals. 
In addition, we express the H-graph with unit powers of the propagators up to weight four in terms of multiple polylogarithms in compact form.
   }
\end{abstract}

\vspace*{\fill}

\newpage

\section{Introduction}
\label{sect:intro}

The observation of gravitational waves~\cite{LIGOScientific:2016aoc} has opened a new research field. Theoretical input in the form of waveform templates is essential to fully exploit experimental facilities. The evolution of a binary system consisting of two massive objects, such as black holes, is typically divided into three stages: the inspiral phase, the merger, and the ringdown.
While the merger and ringdown phases are generally modeled using numerical general relativity, extending this numerical approach to the inspiral phase is theoretically possible but impractical due to the high computational cost. Therefore, analytic methods are preferred for the inspiral phase whenever feasible.

The initial phase of the inspiral process of a binary system producing gravitational waves
can be described by perturbation theory~\cite{Buonanno:1998gg,Buonanno:2000ef,Damour:2012mv,Damour:2017ced,Bini:2020nsb,Bini:2020hmy}
and 
effective field theory methods link general relativity and particle physics~\cite{Goldberger:2004jt, Cheung:2018wkq, Porto:2016pyg, Levi:2018nxp}.
There is a fruitful interplay between gravitational wave physics and techniques developed in the context of perturbative 
quantum field theory~\cite{Foffa:2016rgu,Foffa:2019hrb,Bini:2020uiq,Bini:2020rzn,Bjerrum-Bohr:2018xdl,Cristofoli:2019neg,Kosower:2018adc,Bern:2019nnu,Bern:2019crd,Bern:2021dqo,Bern:2022kto,Bern:2023ccb,Bern:2024adl,Blumlein:2019zku,Blumlein:2019bqq,Blumlein:2020pog,Blumlein:2020znm,Blumlein:2020pyo,Blumlein:2021txj,Foffa:2019rdf,Foffa:2019yfl,Kalin:2019rwq,Kalin:2019inp,Kalin:2020mvi,Kalin:2020fhe,Liu:2021zxr,Dlapa:2021npj,Dlapa:2021vgp,Cho:2022syn,Dlapa:2022lmu,Dlapa:2024cje,Herrmann:2021lqe,DiVecchia:2021bdo,Bjerrum-Bohr:2021vuf,Mogull:2020sak,Jakobsen:2021smu,Jakobsen:2023hig,Jakobsen:2023ndj,Klemm:2024wtd,Driesse:2024xad,Frellesvig:2023bbf,Frellesvig:2024zph}.

At the two-loop level, as discussed in ref.\cite{Bern:2019crd}, the most complex diagram is the so-called H-graph, depicted in fig.~\ref{fig_H-graph}. This graph represents one possible interaction scenario between two massive objects with masses $m_1$ and $m_2$ through graviton exchange. The name H-graph stems from the graviton lines, which form the letter ``H'' (rotated by $90^\circ$ in our figure). In this paper, we compute the H-graph with unequal masses in relativistic quantum field theory.

The simpler equal-mass H-graph has been considered in refs.~\cite{Bianchi:2016yiq, Kreer:2021sdt}. In ref.~\cite{Bianchi:2016yiq}, an $\eps$-factorized differential equation was derived for the master integrals of the equal-mass H-graph, while ref.~\cite{Kreer:2021sdt} expressed these master integrals in terms of multiple polylogarithms.
We further note that in the context of gravitational applications, the  H-graph is only needed within the soft expansion. The associated simplifications were exploited in ref.~\cite{Bern:2019crd} to compute the sum of the unequal-mass H-graph with its crossed counterpart within the soft expansion.

Since this computation relies on subtle limits, a full computation in relativistic quantum field theory provides valuable cross-checks. Moreover, the insights into the structure of these Feynman integrals are helpful at higher loop order, as the most challenging higher-loop diagrams are the ones where the H-graph is dressed up with additional gravitons. We might also say -- following a seminal paper of Steven Weinberg~\cite{Weinberg:1965nx} -- that we compute the H-graph because we can.

We calculate the H-graph with state-of-the-art methods from perturbative quantum field theory:
We use integration-by-parts identities~\cite{Tkachov:1981wb,Chetyrkin:1981qh} to derive the differential equation for the master integrals~\cite{Kotikov:1990kg,Kotikov:1991pm,Remiddi:1997ny,Gehrmann:1999as}.
The essential step of our work is to rotate the basis of master integrals such that 
the differential equation is $\eps$-factorised~\cite{Henn:2013pwa}.
This will introduce six square roots~\cite{Kreer:2021rim}, which cannot be rationalized simultaneously~\cite{Festi:2022sqa}. 
For this reason, we present the result for the master integrals in terms of iterated integrals.
The known equal-mass case provides convenient boundary values for the integration of the differential equation.
The results for the unequal-mass case are then simply given by iterated integrals in the mass ratio $m_2^2/m_1^2$ from
the equal mass value $m_2^2/m_1^2=1$ to the desired value.
However, it is worth noting that the H-graph with unit powers of the propagators can be expressed up to weight four in terms of multiple polylogarithms, and we give a compact expression for this master integral.

This paper is organized as follows: In section~\ref{sect:notation}, we introduce our notation.
The master integrals are defined in section~\ref{sect:masters}. In section~\ref{sect:differential_equation}, we give the $\eps$-factorized differential equation for the master integrals and the corresponding alphabet.
Our results are presented in section~\ref{sect:results}. Finally, our conclusions are given in section~\ref{sect:conclusions}.
The appendices provide additional details:
Appendix~\ref{sect:supplement} describes in detail the electronic file attached to the arXiv version of this article, containing our results in electronic form.
Appendix~\ref{sect:boundary} lists the boundary values used for the integration of the differential equation.
Finally, appendix~\ref{sect:J_28} outlines the calculation of one particular master integral through direct integration in Feynman parameter space.


\section{Notation and setup}
\label{sect:notation}

\begin{figure}
    \begin{center}
    \includegraphics[scale=1.0]{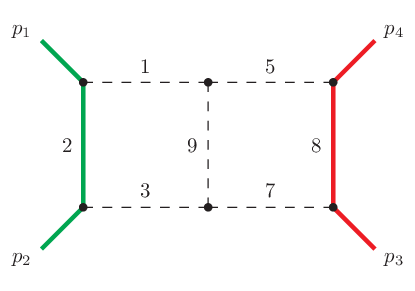}
    \end{center}
    \caption{
    The H-graph.
    The solid green line denotes a massive object of mass $m_1$, 
    the solid red line denotes a massive object of mass $m_2$, dashed lines denote massless particles.
}
\label{fig_H-graph}
\end{figure}

In this section, we clarify our notation and our setup for the computation of the H-graph with unequal masses. Fig.~\ref{fig_H-graph} displays the corresponding Feynman diagram, where the dashed lines form the letter H turned by $90^{\circ}$. Dashed lines correspond to massless particles, while solid green (red) lines represent a massive object of mass $m_1$ ($m_2$). 

\begin{figure}
\begin{center}
\includegraphics[scale=1.0]{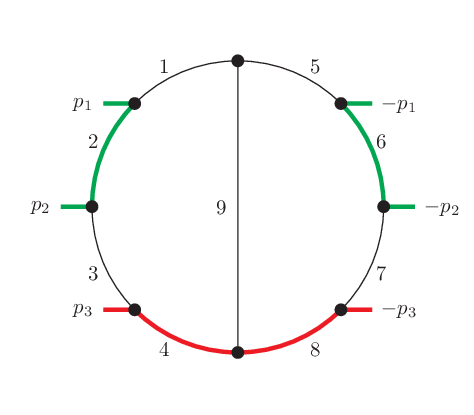}
\end{center}
\caption{
The auxiliary graph.
Green lines correspond to a particle with mass $m_1$, red lines correspond to a particle with mass $m_2$, 
all other lines correspond to massless particles.
}
\label{fig_auxiliary-graph}
\end{figure}

To express any scalar product involving the loop momenta as a linear combination of inverse propagators, we consider an auxiliary graph with nine propagators, shown in Fig.~\ref{fig_auxiliary-graph}. The meaning of dashed, red, and green lines is identical to Fig.~\ref{fig_H-graph}.  

The corresponding family of Feynman integrals is
\bq
\label{def_integral}
 I_{\nu_1 \nu_2 \nu_3 \nu_4 \nu_5 \nu_6 \nu_7 \nu_8 \nu_9}
 & = &
 e^{2 \gamma_E \eps}
 \left(\mu^2\right)^{\nu-D}
 \int \frac{d^Dk_1}{i \pi^{\frac{D}{2}}} \frac{d^Dk_2}{i \pi^{\frac{D}{2}}} 
 \prod\limits_{j=1}^9 \frac{1}{ \left(P_j\right)^{\nu_j} },
\eq
where $D=4-2\eps$ denotes the number of space-time dimensions,
$\gamma_E$ denotes the Euler-Mascheroni constant, 
$\mu$ is an arbitrary scale introduced to render the Feynman integral dimensionless. 
The quantity $\nu$ is defined by
\bq
 \nu & = &
 \sum\limits_{j=1}^9 \nu_j.
\eq
The inverse propagators $P_j$ are defined as follows:
\begin{align}
 P_1 & = -\left(k_1+p_1\right)^2,
 &
 P_2 & = -k_1^2 + m_1^2,
 &
 P_3 & = -\left(k_1-p_2\right)^2,
 \nonumber \\
 P_4 & = -\left(k_1-p_2-p_3\right)^2 + m_2^2,
 &
 P_5 & = -\left(k_2+p_1\right)^2,
 &
 P_6 & = -k_2^2 + m_1^2,
 \nonumber \\
 P_7 & = -\left(k_2-p_2\right)^2,
 & 
 P_8 & = -\left(k_2-p_2-p_3\right)^2 + m_2^2,
 &
 P_9 & = -\left(k_1-k_2\right)^2.
\end{align}
The external momenta satisfy
\bq
 p_1^2 \; = \; p_2^2 \; = \; m_1^2,
 & &
 p_3^2 \; = \; p_4^2 \; = \; m_2^2.
\eq
The Mandelstam variables are defined by
\bq
 s \; = \; \left(p_1+p_2\right)^2, \quad
 t \; = \; \left(p_2+p_3\right)^2, \quad \text{and}\quad
 u \; = \; \left(p_1+p_3\right)^2. 
\eq
In the context of binary sytems, $t$ corresponds to the center of mass energy while $s$ describes the momentum transfer. Hence, $|s| \ll t, m_1^2, m_2^2$ defines the classical limit used in ref.~\cite{Bern:2019crd}. 

The Feynman parameter representation of the integral family in eq.~\eqref{def_integral} is given by
\bq
 I_{\nu_1 \nu_2 \nu_3 \nu_4 \nu_5 \nu_6 \nu_7 \nu_8 \nu_9}
 & = &
 e^{2 \gamma_E \eps}
 \frac{\Gamma(\nu-D)}{\prod\limits_{j=1}^{9}\Gamma(\nu_j)}
 \;
 \int\limits_{a_i \ge 0}
 d^9a
 \;
 \delta\left(1-\sum\limits_{j=1}^9 a_j\right)
 \; 
 \left( \prod\limits_{j=1}^{9} a_j^{\nu_j-1} \right)\,
 \frac{\left[{\mathcal U}\left(a\right)\right]^{\nu-\frac{3}{2}D}}{\left[{\mathcal F}\left(a\right)\right]^{\nu-D}},
 \;\;\;
\eq
where ${\mathcal U}(a)={\mathcal U}(a_1,\dots,a_9)$ and ${\mathcal F}(a)={\mathcal F}(a_1,\dots,a_9)$ 
denote the first and second graph polynomial, respectively. Moreover, we define the derivatives of  ${\mathcal F}$
with respect to 
\bq
 x_k & \in & \left\{\frac{(-s)}{\mu^2},\frac{(-t)}{\mu^2},\frac{m_1^2}{\mu^2},\frac{m_2^2}{\mu^2}\right\}
 \; = \; \left\{ x_{(-s)}, x_{(-t)}, x_{m_1^2}, x_{m_2^2} \right\}
\eq
as 
\bq
    \mathcal{F}_{x_k}' = \mu^2 \frac{d}{dx_k} \mathcal{F}.
\eq
Since ${\mathcal F}$ is linear in the kinematic variables, it satisfies the relation
\bq
 {\mathcal F}
 & = &
 \left( \frac{-s}{\mu^2} \right) {\mathcal F}_{x_{(-s)}}' 
 + \left( \frac{-t}{\mu^2} \right) {\mathcal F}_{x_{(-t)}}' 
 + \left( \frac{m_1^2}{\mu^2} \right) {\mathcal F}_{x_{m_1^2}}'
 + \left( \frac{m_2^2}{\mu^2} \right) {\mathcal F}_{x_{m_2^2}}'.
\eq
We introduce the raising operators ${\bf j}^+$ and the dimensional shift operators ${\bf D}^\pm$ by
\bq
 {\bf j}^+ I_{\nu_1 \dots \nu_j \dots \nu_9}(D)
 & = &
 \nu_j \cdot I_{\nu_1 \dots (\nu_j+1) \dots \nu_9}(D),
 \nonumber \\
 {\bf D}^\pm I_{\nu_1 \dots \nu_9}(D)
 & = &
 I_{\nu_1 \dots \nu_9}(D \pm 2).
\eq
The differential equation for $I_{\nu_1 \dots \nu_9}$
is given by
\bq
 \frac{\partial}{\partial x_k} I_{\nu_1 \dots \nu_9}
 & = &
 - {\mathcal F}_{x_k}'\left({\bf 1}^+,\dots,{\bf 9}^+\right) {\bf D}^+ I_{\nu_1 \dots \nu_9}.
\label{eq:derivative_integral}
\eq

Using the integration-by-parts reduction program {\tt Kira}~\cite{Maierhoefer:2017hyi,Klappert:2020nbg}, we find $40$ master integrals for the family of the H-graph with unequal masses. We denote the pre-canonical master integral basis with $I=(I_1,\dots, I_{40})^T$. We compute the derivative of all master integrals with respect to $x_k$ using eq.~\eqref{eq:derivative_integral}. With the help of integration-by-parts identities~\cite{Tkachov:1981wb, Chetyrkin:1981qh} and
dimensional shift relations~\cite{Tarasov:1996br, Tarasov:1997kx}, we then reduce the integrals on the right-hand side of eq.~\eqref{eq:derivative_integral} to the master integrals $I$. The resulting system of differential equations reads
\bq
 d I\left(x,\eps\right) & = & \tilde{A}\left(x,\eps\right) I\left(x,\eps\right),
\eq
where $x$ denotes the dimensionless kinematic variables. The entries of the $40 \times 40$-matrix $\tilde{A}$ are one-forms, rational in $x$ and $\eps$. 
The essential step in computing the master integrals within the method of differential equations is to find a basis 
$J$ of master integrals, such that the $\eps$-dependence factorizes
and the differential equations become~\cite{Henn:2013pwa}
\bq
 d J\left(x,\eps\right) & = & \eps A\left(x\right) J\left(x,\eps\right),
 \label{eps_factorised_deq}
\eq
where $A$ is now $\eps$-independent. As $I$ and $J$ both are bases of master integrals, they are related by a rotation matrix $U$
\bq
 J\left(x,\eps\right) & = & U\left(x,\eps\right) I\left(x,\eps\right).
\eq
Thus, the differential equations are related by
\bq
 \eps A & = & U \tilde{A} U^{-1} - U d U^{-1}.
\eq

Finding the rotation matrix $U$ is the crucial and non-trivial step in the differential equation method. In general, it is non-rational in $x$ and $\eps$ and may involve square roots and more complicated functions. In our case, it is algebraic and introduces six square roots
\begin{align}
\label{def_roots}
 r_1
 & = 
 \sqrt{-s\vphantom{\left(m_1^2-t\right)^2}} \sqrt{4 m_1^2-s},
 &
 r_4
 & = 
  \sqrt{-s\vphantom{\left(m_1^2-t\right)^2}}\sqrt{4 m_1^2 m_2^4 - s \left( m_1^2-t\right)^2},
 \\
 r_2
 & = 
 \sqrt{-s\vphantom{\left(m_1^2-t\right)^2}} \sqrt{4 m_2^2-s},
 &
 r_5
 & = 
 \sqrt{-s\vphantom{\left(m_1^2-t\right)^2}}\sqrt{4 m_1^4 m_2^2 -s \left( m_2^2 -t \right)^2},
 \nonumber \\
 r_3 
 & =   
 \sqrt{\left(m_1-m_2\right)^2-t} \sqrt{\left(m_1+m_2\right)^2-t},
 &
 r_6 
 & =   
 \sqrt{\left(m_1-m_2\right)^2-s-t} \sqrt{\left(m_1+m_2\right)^2-s-t}.
 \nonumber 
\end{align}
Note that we write
\bq
 \sqrt{-s\vphantom{\left(m_1^2-t\right)^2}} \sqrt{4 m_1^2-s}
 & \mbox{instead of} &
 \sqrt{-s\left(4m_1^2-s\right)}.
\eq
In the Euclidean region ($s<0, t<0, m_1^2>0, m_2^2>0$), the arguments of all roots are positive, and the two forms
are equivalent.
In regions where $s>0$ or $t>0$, we have to add a small imaginary part according to
Feynman's $i\delta$ prescription ($s \rightarrow s+i\delta, t \rightarrow t+i\delta$ with $\delta>0$) and
the two forms may differ (with the standard choice of the branch cut of the square root along the negative real axis).
The form in eq.~\eqref{def_roots} simplifies the analytic continuation from the Euclidean region to the physical
region of interest.


\section{Master integrals}
\label{sect:masters}

In this section, we give a basis $J$ of master integrals, which puts the differential equation into an
$\eps$-factorised form eq.~\eqref{eps_factorised_deq}.
Unless stated otherwise, all integrals are in $D=4-2\eps$ dimensions.
Integrals in $(2-2\eps)$ dimensions are denoted as ${\bf D}^- I_{\nu_1 \dots \nu_9}$.

We derived the 40 master integrals of $J$ from the maximal cuts within the loop-by-loop approach~\cite{Frellesvig:2017aai},
the equal mass limit~\cite{Bianchi:2016yiq,Kreer:2021sdt}, and input from families sharing common sub-topologies of Feynman integrals~\cite{Chaubey:2019lum}.

As master integrals, we take
\bq
\label{def_master_integrals}
 J_{1}
 & = &
 \eps^2 I_{020000020},
 \nonumber \\
 J_{2}
 & = &
 \eps^2 \frac{\left(-s\right)}{\mu^2} I_{020020100},
 \nonumber \\
 J_{3}
 & = &
 \eps^2 \frac{\left(-s\right)}{\mu^2} I_{201000020},
 \nonumber \\
 J_{4}
 & = &
 \frac{\eps^2 \left(1+4\eps\right)}{1+\eps} \frac{m_1^2}{\mu^2} I_{010020002},
 \nonumber \\
 J_{5}
 & = &
 \eps^2 \frac{\left(-s\right)}{\mu^2} I_{001020002},
 \nonumber \\
 J_{6}
 & = &
 \frac{\eps^2 \left(1+4\eps\right)}{1+\eps} \frac{m_2^2}{\mu^2} I_{200000012},
 \nonumber \\
 J_{7}
 & = &
 \eps^2 \frac{r_3}{\mu^2} {\bf D}^- I_{010000011},
 \nonumber \\
 J_{8}
 & = &
 \eps^2 
 \frac{\mu^2}{t+m_1^2-m_2^2} 
  \left[ 
        2 \frac{t}{\mu^2} {\bf D}^- I_{\left(-1\right)10000011} 
        - \frac{r_3^2}{\mu^4} {\bf D}^- I_{010000011} 
  \right],
 \nonumber \\
 J_{9}
 & = &
 \eps^2 
 \frac{\mu^2}{t-m_1^2+m_2^2}
  \left[
        2 \frac{t}{\mu^2} {\bf D}^- I_{0100\left(-1\right)0011}
        - \frac{r_3^2}{\mu^4} {\bf D}^- I_{010000011} 
  \right],
 \nonumber \\
 J_{10}
 & = &
 \eps^2 \frac{\left(-s\right)^2}{\mu^4} I_{201020100},
 \nonumber \\
 J_{11}
 & = &
 \eps^3 \frac{r_1}{\mu^2} I_{111000020},
 \nonumber \\
 J_{12}
 & = &
 \eps^3 \frac{r_2}{\mu^2} I_{020010110},
 \nonumber \\
 J_{13}
 & = &
 2 \eps^3 \frac{r_1}{\mu^2} I_{011020001},
 \nonumber \\
 J_{14}
 & = &
 \eps^3 \frac{r_1}{\mu^2} I_{020010101},
 \nonumber \\
 J_{15}
 & = &
 \eps^2 \frac{m_1^2 r_1}{\mu^4} I_{030010101},
 \nonumber \\
 J_{16}
 & = &
 \eps^2 \frac{\left(-s\right)}{\mu^2}
 \left[ 
        \frac{3}{2} \eps I_{020010101} + \frac{m_1^2}{\mu^2} I_{020020101} - \frac{m_1^2}{\mu^2} I_{030010101}
 \right],
 \nonumber \\
 J_{17}
 & = &
 2 \eps^3 \frac{r_3}{\mu^2} I_{110000021},
 \nonumber \\
 J_{18}
 & = &
 \eps^3 \frac{r_2}{\mu^2} I_{101000021},
 \nonumber \\
 J_{19}
 & = &
 \eps^2 \frac{m_2^2 r_2}{\mu^4} I_{101000031},
 \nonumber \\
 J_{20}
 & = &
 \eps^2 \frac{\left(-s\right)}{\mu^2}
 \left[ 
        \frac{3}{2} \eps I_{101000021} + \frac{m_2^2}{\mu^2} I_{201000021} - \frac{m_2^2}{\mu^2} I_{101000031}
 \right],
 \nonumber \\
 J_{21}
 & = &
 2 \eps^3 \frac{r_3}{\mu^2} I_{020010011},
 \nonumber \\
 J_{22}
 & = &
 2 \eps^3 \frac{r_2}{\mu^2} I_{002010011},
 \nonumber \\
 J_{23}
 & = &
 \eps^3 \frac{\left(-s\right) r_1}{\mu^4} I_{111020100},
 \nonumber \\
 J_{24}
 & = &
 \eps^3 \frac{\left(-s\right) r_2}{\mu^4} I_{201010110},
 \nonumber \\
 J_{25}
 & = &
 \eps^3 \frac{r_4}{\mu^4} I_{111000021},
 \nonumber \\
 J_{26}
 & = &
 \eps^3 \frac{r_1}{\mu^2} \left( I_{111\left(-1\right)00021} - \frac{m_2^2}{\mu^2} I_{111000021} \right),
 \nonumber \\
 J_{27}
 & = &
 \eps^2 \frac{\left(-s\right) r_3}{\mu^4} \left( \frac{m_2^2}{\mu^2} I_{111000031} - \eps I_{111000021} \right),
 \nonumber \\
 J_{28}
 & = &
 4 \eps^4 \frac{r_6}{\mu^2} I_{011010011},
 \nonumber \\
 J_{29}
 & = &
 2 \eps^3 \frac{\left(-s\right) r_3}{\mu^4} I_{011010012},
 \nonumber \\
 J_{30}
 & = &
 2 \eps^3 \frac{m_2^2 r_1}{\mu^4} I_{011010021},
 \nonumber \\
 J_{31}
 & = &
 2 \eps^3 \frac{m_1^2 r_2}{\mu^4} I_{021010011},
 \nonumber \\
 J_{32}
 & = &
 \frac{1}{2} \eps^2 \frac{\left(-s\right)}{\mu^2} 
  \left[
   2 \frac{m_1^2 m_2^2}{\mu^4} I_{021010021} - 2 \eps \frac{m_2^2}{\mu^2} I_{011010021} - 2 \eps \frac{m_1^2}{\mu^2} I_{021010011} 
 \right. \nonumber \\
 & & \left.
   - \eps \frac{\left( m_1^2 + m_2^2 - t \right)}{\mu^2} I_{011010012}
  \right],
 \nonumber \\
 J_{33}
 & = &
 \eps^3 \frac{r_5}{\mu^4} I_{020010111},
 \nonumber \\
 J_{34}
 & = &
 \eps^3 \frac{r_2}{\mu^2} \left( I_{02001\left(-1\right)111} - \frac{m_1^2}{\mu^2} I_{020010111} \right),
 \nonumber \\
 J_{35}
 & = &
 \eps^2 \frac{\left(-s\right) r_3}{\mu^4} \left( \frac{m_1^2}{\mu^2} I_{030010111} - \eps I_{020010111} \right),
 \nonumber \\
 J_{36}
 & = &
 \eps^4 \frac{r_1 r_2}{\mu^4} I_{111010110},
 \nonumber \\
 J_{37}
 & = &
 \eps^4 \frac{\left(-s\right)^2 r_3}{\mu^6} I_{111010111},
 \nonumber \\
 J_{38}
 & = &
 \eps^4 \frac{\left(-s\right) r_1}{\mu^4} I_{111\left(-1\right)10111},
 \nonumber \\
 J_{39}
 & = &
 \eps^4 \frac{\left(-s\right) r_2}{\mu^4} I_{11101\left(-1\right)111},
 \nonumber \\
 J_{40}
 & = &
 \eps^4 \frac{\left(-s\right)}{\mu^2}
 \left[
       I_{111\left(-1\right)1\left(-1\right)111} 
       + \frac{1}{2} \frac{\left(-s\right)}{\mu^2} I_{111\left(-1\right)10111}
       + \frac{1}{2} \frac{\left(-s\right)}{\mu^2} I_{11101\left(-1\right)111}
 \right]
 \nonumber \\
 & &
 - \eps^4 \frac{\left[
                \left(-s\right)\left(-t\right) + \left(-s\right) \left(m_1^2+m_2^2-\frac{1}{2} s \right) - \frac{1}{2} r_1 r_2
          \right]}{\mu^4} I_{111010110}
 \nonumber \\
 & &
 + \eps^3 \frac{\left(-s\right)}{\mu^2} 
   \left[
     -2 \eps I_{011010011}
     + \frac{m_2^2}{\mu^2} I_{011010021}
     + \frac{m_1^2}{\mu^2} I_{021010011}
     + I_{111\left(-1\right)00021} - \frac{m_2^2}{\mu^2} I_{111000021}
 \right. \nonumber \\
 & & \left.
     + I_{02001\left(-1\right)111} - \frac{m_1^2}{\mu^2} I_{020010111}
     + \frac{1}{2} \frac{\left(-s\right)}{\mu^2} I_{111020100}
     + \frac{1}{2} \frac{\left(-s\right)}{\mu^2} I_{201010110}
     + \frac{1}{2} I_{020010101}
 \right. \nonumber \\
 & & \left.
     + \frac{1}{2} I_{101000021}
     - \frac{1}{\eps} \frac{m_1^2}{\mu^2} I_{030010101}
     - \frac{1}{\eps} \frac{m_2^2}{\mu^2} I_{101000031}
     - 2 I_{011020001}
     - 2 I_{002010011}
 \right. \nonumber \\
 & & \left.
     - \frac{1}{2} I_{111000020}
     - \frac{1}{2} I_{020010110}
   \right]
 + \frac{\eps^2}{1-2\eps} \frac{\left(-s\right)}{\mu^2}
   \left[
          - \eps \frac{\left(-s\right)}{\mu^2} I_{201020100}
          + \frac{1}{2} I_{001020002}
   \right].
\eq


\section{The differential equation and differential forms}
\label{sect:differential_equation}

The differential equation for the basis of master integrals $J=(J_1,\dots,J_{40})^T$ is in $\eps$-factorised form
\bq
\label{diff_eq_J}
 d J\left(x,\eps\right) & = & \eps A\left(x\right) J\left(x,\eps\right).
\eq
We write
\bq
 A & = &
 \sum\limits_{k=1}^{\Nletter} C_k \omega_k
 \;\;\;\;\;\;
 \mbox{with} \;\; \Nletter \; = \; 29,
\eq
where the $C_k$s are $40 \times 40$-matrices, whose entries are rational numbers. We find $\Nletter=29$ differential one-forms $\omega_k$. We call the $\omega_k$s letters and the set $\{\omega_1,\dots,\omega_{29}\}$ the alphabet. We sort the letters by their square root dependence. The letters without roots are
\bq
\label{def_differential_one_forms}
 \omega_{1}
 & = & 
 d \ln\left(\frac{-s}{\mu^2}\right),
 \nonumber \\
 \omega_{2}
 & = & 
 d \ln\left(\frac{-t}{\mu^2}\right),
 \nonumber \\
 \omega_{3}
 & = & 
 d \ln\left(\frac{m_1^2}{\mu^2}\right),
 \nonumber \\
 \omega_{4}
 & = & 
 d \ln\left(\frac{m_2^2}{\mu^2}\right),
 \nonumber \\
 \omega_{5}
 & = & 
 d \ln\left(\frac{4m_1^2-s}{\mu^2}\right),
 \nonumber \\
 \omega_{6}
 & = & 
 d \ln\left(\frac{4m_2^2-s}{\mu^2}\right),
 \nonumber \\
 \omega_{7}
 & = & 
 d \ln\left(\frac{\left(m_1^2-m_2^2\right)^2-2\left(m_1^2+m_2^2\right)t+t^2}{\mu^4}\right),
 \nonumber \\
 \omega_{8}
 & = & 
 d \ln\left(\frac{\left(m_1^2-m_2^2\right)^2-2\left(m_1^2+m_2^2\right)t+t^2+st}{\mu^4}\right),
 \nonumber \\
 \omega_{9}
 & = & 
 d \ln\left(\frac{\left(m_1^2-m_2^2\right)^2-2\left(m_1^2+m_2^2\right)\left(s+t\right)+\left(s+t\right)^2}{\mu^4}\right).
\eq
The letters involving one root are
\bq
 \omega_{10}
 & = & 
 d \ln\left(\frac{2m_1^2-s-r_1}{2m_1^2-s+r_1}\right),
 \nonumber \\
 \omega_{11}
 & = & 
 d \ln\left(\frac{2m_2^2-s-r_2}{2m_2^2-s+r_2}\right),
 \nonumber \\
 \omega_{12}
 & = & 
 d \ln\left(\frac{m_1^2+m_2^2-t-r_3}{m_1^2+m_2^2-t+r_3}\right),
 \nonumber \\
 \omega_{13}
 & = & 
 d \ln\left(\frac{\left(m_1^2-m_2^2\right)^2-\left(m_1^2+m_2^2\right)t-\left(m_1^2-m_2^2\right) r_3}{\left(m_1^2-m_2^2\right)^2-\left(m_1^2+m_2^2\right)t+\left(m_1^2-m_2^2\right) r_3}\right),
 \nonumber \\
 \omega_{14}
 & = & 
 d \ln\left(\frac{\left(-s\right)\left(m_1^2-t\right)-r_4}{\left(-s\right)\left(m_1^2-t\right)+r_4}\right),
 \nonumber \\
 \omega_{15}
 & = & 
 d \ln\left(\frac{\left(-s\right)\left(m_2^2-t\right)-r_5}{\left(-s\right)\left(m_2^2-t\right)+r_5}\right),
 \nonumber \\
 \omega_{16}
 & = & 
 d \ln\left(\frac{m_1^2+m_2^2-s-t-r_6}{m_1^2+m_2^2-s-t+r_6}\right).
\eq
The letters with two roots are
\bq
\label{def_differential_one_forms_III}
 \omega_{17}
 & = & 
 d \ln\left(\frac{\left(-s\right)\left(2\left(m_1^2+m_2^2-t\right)-s\right)-r_1 r_2}{\left(-s\right)\left(2\left(m_1^2+m_2^2-t\right)-s\right)+r_1 r_2}\right),
 \nonumber \\
 \omega_{18}
 & = & 
 d \ln\left(\frac{\left(-s\right)\left(m_2^2-m_1^2-t\right)-r_1 r_3}{\left(-s\right)\left(m_2^2-m_1^2-t\right)+r_1 r_3}\right),
 \nonumber \\
 \omega_{19}
 & = & 
 d \ln\left(\frac{\left(-s\right)\left(m_1^2-m_2^2-t\right)-r_2 r_3}{\left(-s\right)\left(m_1^2-m_2^2-t\right)+r_2 r_3}\right),
 \nonumber \\
 \omega_{20}
 & = & 
 d \ln\left(\frac{\left(-s\right)\left(4 m_1^2 m_2^2 - m_1^2s-st\right)-r_1 r_4}{\left(-s\right)\left(4 m_1^2 m_2^2 - m_1^2s-st\right)+r_1 r_4}\right),
 \nonumber \\
 \omega_{21}
 & = & 
 d \ln\left(\frac{p_{21} - q_{21} r_2 r_4}{p_{21} + q_{21} r_2 r_4}\right),
 \nonumber \\
 \omega_{22}
 & = & 
 d \ln\left(\frac{\left(-s\right)\left(\left(m_1^2-t\right)^2-m_2^2\left(m_1^2+t\right)\right)-r_3 r_4}{\left(-s\right)\left(\left(m_1^2-t\right)^2-m_2^2\left(m_1^2+t\right)\right)+r_3 r_4}\right),
 \nonumber \\
 \omega_{23}
 & = & 
 d \ln\left(\frac{p_{23} - q_{23} r_1 r_5}{p_{23} + q_{23} r_1 r_5}\right),
 \nonumber \\
 \omega_{24}
 & = & 
 d \ln\left(\frac{\left(-s\right)\left(4 m_1^2 m_2^2 - m_2^2s-st\right)-r_2 r_5}{\left(-s\right)\left(4 m_1^2 m_2^2 - m_2^2s-st\right)+r_2 r_5}\right),
 \nonumber \\
 \omega_{25}
 & = & 
 d \ln\left(\frac{\left(-s\right)\left(\left(m_2^2-t\right)^2-m_1^2\left(m_2^2+t\right)\right)-r_3 r_5}{\left(-s\right)\left(\left(m_2^2-t\right)^2-m_1^2\left(m_2^2+t\right)\right)+r_3 r_5}\right),
 \nonumber \\
 \omega_{26}
 & = & 
 d \ln\left(\frac{p_{26} - q_{26} r_1 r_6}{p_{26} + q_{26} r_1 r_6}\right),
 \nonumber \\
 \omega_{27}
 & = & 
 d \ln\left(\frac{p_{27} - q_{27} r_2 r_6}{p_{27} + q_{27} r_2 r_6}\right),
 \nonumber \\
 \omega_{28}
 & = & 
 d \ln\left(\frac{\left(m_1^2-m_2^2\right)^2-\left(m_1^2+m_2^2\right)\left(s+2t\right)+st+t^2 - r_3 r_6}{\left(m_1^2-m_2^2\right)^2-\left(m_1^2+m_2^2\right)\left(s+2t\right)+st+t^2 + r_3 r_6}\right),
 \nonumber \\
 \omega_{29}
 & = & 
 d \ln\left(\frac{p_{29} - q_{29} r_3 r_6}{p_{29} + q_{29} r_3 r_6}\right),
\eq
with the polynomials
\bq
 p_{21}
 & = &
 \left(-s\right) 
   \left(
          2 m_2^4 \left( \left(m_1^2+m_2^2\right)^2 + 4 m_1^2 m_2^2 - 2 \left(m_1^2+m_2^2\right)t + t^2 \right) 
 \right. \nonumber \\
 & & \left.
          + 2 m_2^2 \left(-s\right) \left( 2 m_1^4 + 2 m_1^2 m_2^2 - 4m_1^2 t- m_2^2 t + 2 t^2 \right)
          + \left(-s\right)^2 \left(m_1^2-t\right)^2
   \right),
 \nonumber \\
 q_{21}
 & = &
 2 m_2^2\left(m_1^2+m_2^2-t\right) + \left(-s\right) \left(m_1^2-t\right),
 \nonumber \\
 p_{23}
 & = &
 \left(-s\right) 
   \left(
          2 m_1^4 \left( \left(m_1^2+m_2^2\right)^2 + 4 m_1^2 m_2^2 - 2 \left(m_1^2+m_2^2\right)t + t^2 \right) 
 \right. \nonumber \\
 & & \left.
          + 2 m_1^2 \left(-s\right) \left( 2 m_2^4 + 2 m_1^2 m_2^2 - 4m_2^2 t- m_1^2 t + 2 t^2 \right)
          + \left(-s\right)^2 \left(m_2^2-t\right)^2
   \right),
 \nonumber \\
 q_{23}
 & = &
 2 m_1^2\left(m_1^2+m_2^2-t\right) + \left(-s\right) \left(m_2^2-t\right),
 \nonumber \\
 p_{26}
 & = &
 2 m_1^2 \left(m_1^2-m_2^2\right)^2 - 4 m_1^2\left(m_1^2+m_2^2\right) t + 2m_1^2 t^2 + 2 \left(4m_1^2+m_2^2\right) s t 
 \nonumber \\
 & &
 - s \left(3m_1^2+m_2^2\right)\left(3m_1^2+m_2^2-2s\right) - s \left(s+t\right)^2,
 \nonumber \\
 q_{26}
 & = &
 3 m_1^2 + m_2^2 -s -t,
 \nonumber \\
 p_{27}
 & = &
 2 m_2^2 \left(m_1^2-m_2^2\right)^2 - 4 m_2^2\left(m_1^2+m_2^2\right) t + 2m_2^2 t^2 + 2 \left(4m_2^2+m_1^2\right) s t 
 \nonumber \\
 & &
 - s \left(3m_2^2+m_1^2\right)\left(3m_2^2+m_1^2-2s\right) - s \left(s+t\right)^2,
 \nonumber \\
 q_{27}
 & = &
 3 m_2^2 + m_1^2 -s -t,
 \nonumber \\
 p_{29}
 & = &
 \left(m_1^2-m_2^2\right)^2\left(m_1^2+m_2^2-s\right) + \left(-t\right) \left(m_1^2+m_2^2\right) \left(2m_1^2+2m_2^2-s-t\right),
 \nonumber \\
 q_{29}
 & = &
 m_1^2-m_2^2.
\eq
We provide the matrix $A$ in electronic form, see appendix~\ref{sect:supplement}.


\section{The solution in terms of iterated integrals}
\label{sect:results}

Without loss of generality, we set in the following
\bq
 \mu^2 & = & m_1^2.
\eq
The $\eps$-factorized differential equation eq.~\eqref{diff_eq_J} is easily solved
in terms of iterated integrals. More precisely, the expansion coefficients $J_i^{(j)}$ in
\bq
 J_i 
 & = & 
 \sum\limits_{j=0}^\infty J_i^{(j)} \; \eps^j,
 \;\;\;\;\;\;
 i \in \{1,\dots,40\}
\eq
are  linear combinations of iterated integrals. 

An iterated integral is defined as follows~\cite{Chen}:
Let $M$ be a $n$-dimensional manifold
and
\bq
 \gamma
 & : &
 \left[a,b\right] \rightarrow M
\eq
a path with start point ${x}_i=\gamma(a)$ and end point ${x}_f=\gamma(b)$.
Suppose further that $\omega_1$, ..., $\omega_r$ are differential one-forms on $M$.
Let us write
\bq
 f_j\left(\lambda\right) d\lambda & = & \gamma^\ast \omega_j
\eq
for the pull-backs to the interval $[a,b]$.
For $\lambda \in [a,b]$ and $f_r(\lambda)$ regular at $\lambda=0$ the $r$-fold iterated integral of $\omega_1$, \dots, $\omega_r$
along the path $\gamma$ is defined by
\bq
 I_{\gamma}\left(\omega_1,...,\omega_r;\lambda\right)
 & = &
 \int\limits_a^{\lambda} d\lambda_1 f_1\left(\lambda_1\right)
 \int\limits_a^{\lambda_1} d\lambda_2 f_2\left(\lambda_2\right)
\dots
 \int\limits_a^{\lambda_{r-1}} d\lambda_r f_r\left(\lambda_r\right).
\eq
We regulate potential simple poles at $\lambda=0$ in $f_r(\lambda)$ with the standard ``trailing zero'' or tangential base point
prescription, see for example refs.~\cite{Brown:2013qva,Brown:2014aa,Walden:2020odh}.

The six square roots $r_1$-$r_6$ are not simultaneously rationalizable \cite{Festi:2022sqa}.
This does not imply that the master integrals cannot be expressed in terms of multiple polylogarithms \cite{Besier:2019hqd,Heller:2019gkq}.
However, in view of efficiency and given the size of the alphabet and the number of square roots, it is not advisable
to try to express them in terms of multiple polylogarithms.
Not every master integral involves all square roots, and by rationalizing the occurring square roots with
the techniques of refs.~\cite{Besier:2018jen,Besier:2019kco} we were able to express some master integrals in terms of multiple polylogarithms.
However, this will produce expressions with a huge number of terms.
The results can be presented in a more compact form by staying within the iterated integrals.

For the integration of the differential equation, we need boundary data. It is straightforward to impose the boundary condition on the hyper-surface $m_1^2=m_2^2$ for arbitrary $s$ and $t$ from the equal-mass case~\cite{Bianchi:2016yiq, Kreer:2021sdt}. 
In appendix~\ref{sect:boundary}, we summarise how the master integrals $J_1$-$J_{40}$ of the unequal-mass
H-graph degenerate to the master integrals $J_1^{\mathrm{eq}}$-$J_{25}^{\mathrm{eq}}$ of the equal-mass H-graph
in the equal-mass limit. Hence, we only need to integrate the differential equation in one kinematic variable,
which, for convenience, we take as
\bq
 z & = & \frac{m_1-m_2}{m_1}.
\eq
The equal mass case corresponds to $z=0$. Thus, we integrate from $z=0$ to any desired value $z$. This defines the integration path $\gamma$.

The result in terms of iterated integrals is provided in electronic form in a file attached to the arXiv version of this article. 
Details of this file are described in appendix~\ref{sect:supplement}.
The iterated integrals can be evaluated numerically with the help 
of the class \verb|user_defined_kernel| within \verb|GiNaC|~\cite{Walden:2020odh}.

The region
\bq
 s<0,
 \;\;\;\;\;\;\;\;\;
 t>(m_1+m_2)^2,
 \;\;\;\;\;\;\;\;\;
 u<0
\eq
is relevant to the inspiral process of a binary system.
Table~\ref{table_numerical_results} provides reference values for all master integrals for the first five terms of the $\eps$-expansion at the kinematic point
\bq
\label{kinematic_point}
 s \; = \; - \frac{1}{36} \; \mathrm{GeV}^2,
 \;\;\;\;\;\;
 t \; = \; 5 \; \mathrm{GeV}^2,
 \;\;\;\;\;\;
 m_1^2 \; = \; 1 \; \mathrm{GeV}^2,
 \;\;\;\;\;\;
 m_2^2 \; = \; \frac{3}{4} \; \mathrm{GeV}^2.
\eq
We recall that we set $\mu^2=m_1^2$.
\begin{table}[!htbp]
\begin{center}
{\scriptsize
\begin{tabular}{|l|lllll|}
 \hline 
 & $\eps^0$ & $\eps^1$ & $\eps^2$ & $\eps^3$ & $\eps^4$ \\
 \hline 
 $J_{ 1}$ & $        1$ & $ 0.28768207$ & $ 1.6863146$ & $-0.32418509$ & $ 1.7318791$ \\ 
$J_{ 2}$ & $       -1$ & $-3.5835189$ & $-6.420804$ & $-4.4642058$ & $ 7.8627655$ \\ 
$J_{ 3}$ & $       -1$ & $-3.871201$ & $-7.4930986$ & $-6.4636119$ & $ 6.2982922$ \\ 
$J_{ 4}$ & $     -0.5$ & $        0$ & $-4.1123352$ & $-4.407542$ & $-40.992992$ \\ 
$J_{ 5}$ & $        1$ & $ 7.1670379$ & $ 24.038282$ & $ 36.746282$ & $-39.627704$ \\ 
$J_{ 6}$ & $     -0.5$ & $-0.28768207$ & $-4.1950961$ & $-6.7895048$ & $-44.211899$ \\ 
$J_{ 7}$ & $        0$ & $-2.4849066 +  6.2831853 i$ & $-43.839189 -19.719222 i$ & $ 17.022982 -132.5544 i$ & $ 224.25107 -178.24249 i$ \\ 
$J_{ 8}$ & $        0$ & $ 0.28768207$ & $-8.4615925 -7.3082921 i$ & $ 19.880219 -26.261312 i$ & $ 38.536755 -22.48578 i$ \\ 
$J_{ 9}$ & $        0$ & $-0.28768207$ & $-8.2316163 -8.3048369 i$ & $ 18.666637 -29.863999 i$ & $ 52.640226 -25.655134 i$ \\ 
$J_{ 10}$ & $        1$ & $ 7.1670379$ & $ 24.038282$ & $ 43.958624$ & $ 21.804329$ \\ 
$J_{ 11}$ & $        0$ & $        0$ & $ 8.9398354$ & $ 51.556944$ & $  182.634$ \\ 
$J_{ 12}$ & $        0$ & $        0$ & $ 8.8515924$ & $ 48.883259$ & $ 168.15532$ \\ 
$J_{ 13}$ & $        0$ & $        0$ & $-17.879671$ & $-168.99678$ & $-962.19027$ \\ 
$J_{ 14}$ & $        0$ & $        0$ & $        0$ & $ 3.4772053$ & $ 19.173012$ \\ 
$J_{ 15}$ & $        0$ & $        0$ & $ 4.4699177$ & $ 45.726399$ & $ 286.64965$ \\ 
$J_{ 16}$ & $     0.25$ & $        0$ & $-2.7991685$ & $-3.7744625$ & $ 40.39835$ \\ 
$J_{ 17}$ & $        0$ & $        0$ & $        0$ & $-31.464401 -20.636744 i$ & $-96.529606 -162.21025 i$ \\ 
$J_{ 18}$ & $        0$ & $        0$ & $        0$ & $ 3.7133903$ & $ 21.559176$ \\ 
$J_{ 19}$ & $        0$ & $        0$ & $ 4.4257962$ & $ 45.871639$ & $ 288.53509$ \\ 
$J_{ 20}$ & $     0.25$ & $ 0.14384104$ & $-2.2630212$ & $-2.6605466$ & $ 42.356753$ \\ 
$J_{ 21}$ & $        0$ & $        0$ & $        0$ & $-28.821152 -18.160424 i$ & $-85.304855 -142.16077 i$ \\ 
$J_{ 22}$ & $        0$ & $        0$ & $-17.703185$ & $ -168.633$ & $-961.92304$ \\ 
$J_{ 23}$ & $        0$ & $        0$ & $-8.9398354$ & $-81.021183$ & $-386.4065$ \\ 
$J_{ 24}$ & $        0$ & $        0$ & $-8.8515924$ & $-80.603108$ & $-385.60345$ \\ 
$J_{ 25}$ & $        0$ & $        0$ & $        0$ & $ 25.938682 +  24.140129 i$ & $ 217.26845 +  280.98972 i$ \\ 
$J_{ 26}$ & $        0$ & $        0$ & $        0$ & $-25.385867 -25.596324 i$ & $-206.0294 -285.96052 i$ \\ 
$J_{ 27}$ & $        0$ & $-0.62122666 +  1.5707963 i$ & $-7.9435211 +  7.6280288 i$ & $-25.606664 +  10.188395 i$ & $ 25.445655 -22.515026 i$ \\ 
$J_{ 28}$ & $        0$ & $        0$ & $        0$ & $        0$ & $-295.10478 -99.913704 i$ \\ 
$J_{ 29}$ & $        0$ & $ 2.4849066 -6.2831853 i$ & $ 31.774084 -30.512115 i$ & $ 132.56943 -21.354996 i$ & $ 48.678998 +  253.68516 i$ \\ 
$J_{ 30}$ & $        0$ & $        0$ & $-8.9398354$ & $-66.668098 +  51.192649 i$ & $-270.50816 +  632.72706 i$ \\ 
$J_{ 31}$ & $        0$ & $        0$ & $-8.8515924$ & $-66.316811 +  51.037204 i$ & $-271.34154 +  630.80279 i$ \\ 
$J_{ 32}$ & $      0.5$ & $ 3.5835189$ & $ 9.5011179 -3.9032822 i$ & $ 12.583873 -28.025573 i$ & $ 92.157077 -54.070322 i$ \\ 
$J_{ 33}$ & $        0$ & $        0$ & $        0$ & $ 26.971856 +  24.156794 i$ & $ 222.16842 +  280.10239 i$ \\ 
$J_{ 34}$ & $        0$ & $        0$ & $        0$ & $-26.716463 -25.518602 i$ & $-213.42181 -285.38339 i$ \\ 
$J_{ 35}$ & $        0$ & $-0.62122666 +  1.5707963 i$ & $-7.9435211 +  7.6280288 i$ & $-25.606664 +  10.188395 i$ & $ 24.667114 -22.722722 i$ \\ 
$J_{ 36}$ & $        0$ & $        0$ & $        0$ & $        0$ & $ 79.131779$ \\ 
$J_{ 37}$ & $        0$ & $        0$ & $        0$ & $ 15.071388 +  9.699292 i$ & $ 123.10196 +  92.180609 i$ \\ 
$J_{ 38}$ & $        0$ & $        0$ & $        0$ & $        0$ & $-52.944338 -40.123472 i$ \\ 
$J_{ 39}$ & $        0$ & $        0$ & $        0$ & $        0$ & $-52.890947 -39.998028 i$ \\ 
$J_{ 40}$ & $      0.5$ & $ 3.5835189$ & $ 12.019141$ & $ 18.373141$ & $ 106.70776 +  17.304891 i$ \\ 
 \hline 
\end{tabular}
}
\end{center}
\caption{
Numerical results for the first five terms of the $\eps$-expansion of the master integrals $J_{1}$-$J_{40}$ 
at the kinematic point of eq.~\eqref{kinematic_point}.
}
\label{table_numerical_results}
\end{table}
The values in table~\ref{table_numerical_results} are given to $8$ digits precision.
A higher precision can easily be reached.
We have numerically verified our results with the help of the program \verb|AMFlow| \cite{Liu:2022chg}.

The master integrals $J_{28}$ and $J_{37}$ deserve to be discussed in detail.
Up to order $\eps^4$ the square root $r_6$ enters only the master integral $J_{28}$.
This integral can be expressed in term multiple polylogarithms by direct integration in Feynman parameter space, following the lines
of ref.~\cite{Heller:2019gkq}.
The Feynman integral $J_{28}$ starts at ${\mathcal O}(\eps^4)$ and as we are only interested in terms up to weight four, 
we only need to compute $J_{28}^{(4)}$.
We present the details in appendix~\ref{sect:J_28}.

The master integral $J_{37}$ is proportional to the H-graph integral $I_{111010111}$. Up to weight four, its multiple polylogarithm representation is rather simple:
\bq
\label{eq_J37}
\lefteqn{
 J_{37}
 = 
 \left\{
 \frac{1}{2}  I_{\gamma}\left(\omega_{12},\omega_{12},\omega_{12};z\right)
 + G\left(1;\bar{y}\right) I_{\gamma}\left(\omega_{12},\omega_{12};z\right)
 + 2 \left[ G\left(1,1;\bar{y}\right) +\zeta_2 \right] I_{\gamma}\left(\omega_{12};z\right)
 \right. } \nonumber \\
 & & \left.
 + 4 \left[ G\left(1,1,1;\bar{y}\right) + \zeta_2 G\left(1;\bar{y}\right) \right] \right\} \eps^3
 + \left\{
           \frac{1}{2}  I_{\gamma}\left(\omega_7-\omega_4,\omega_{12},\omega_{12},\omega_{12};z\right)
 \right. \nonumber \\
 & & \left.
           - \frac{1}{2}  I_{\gamma}\left(\omega_{12},\omega_{12},\omega_7-\omega_4,\omega_{12};z\right)
           + \left[ G\left(1,\bar{x}\right) -2 \ln\left(\bar{x}\right) \right] I_{\gamma}\left(\omega_{12},\omega_{12},\omega_{12};z\right)
 \right. \nonumber \\
 & & \left.
           + G\left(1;\bar{y}\right) 
             \left[ I_{\gamma}\left(\omega_7-\omega_4,\omega_{12},\omega_{12};z\right) - I_{\gamma}\left(\omega_{12},\omega_{12},\omega_7-\omega_4;z\right) \right]
 \right. \nonumber \\
 & & \left.
           + 2 \left[ G\left(1,1;\bar{y}\right)-G\left(2,1;\bar{y}\right)-G\left(0,1;\bar{y}\right)
                      +G\left(1;\bar{y}\right) G\left(1;\bar{x}\right)-2G\left(1;\bar{y}\right) \ln\left(\bar{x}\right) 
               \right] I_{\gamma}\left(\omega_{12},\omega_{12};z\right)
 \right. \nonumber \\
 & & \left.
           +2 \left[ G\left(1,1;\bar{y}\right) + \zeta_2 \right] I_{\gamma}\left(\omega_7-\omega_4,\omega_{12};z\right)
           + \left[ 4 G\left(1,1,1;\bar{y}\right) - 4 G\left(1,2,1;\bar{y}\right) - 4 G\left(1,0,1;\bar{y}\right)
 \right. \right. \nonumber \\
 & & \left. \left.
                    + 4 G\left(1,1;\bar{y}\right) G\left(1;\bar{x}\right) + 4 \zeta_2 G\left(1;\bar{x}\right)
                    - 8 G\left(1,1;\bar{y}\right) \ln\left(\bar{x}\right) - 8 \zeta_2 \ln\left(\bar{x}\right)
                    - 2 \zeta_3 
             \right] I_{\gamma}\left(\omega_{12};z\right)
 \right. \nonumber \\
 & & \left.
           + 4 \left[ G\left(1,1,1;\bar{y}\right) + \zeta_2 G\left(1;\bar{y}\right) \right] I_{\gamma}\left(\omega_7-\omega_4;z\right)
           + 8 G\left(2,1,1,1;\bar{y}\right)
           + 8 G\left(0,1,1,1;\bar{y}\right)
 \right. \nonumber \\
 & & \left.
           - 8 G\left(1,1,2,1;\bar{y}\right)
           - 8 G\left(1,1,0,1;\bar{y}\right)
           + 8 \zeta_2 \left[ G\left(2,1;\bar{y}\right) + G\left(0,1;\bar{y}\right) - G\left(1,1;\bar{y}\right) \right]
 \right. \nonumber \\
 & & \left.
           - 4 \zeta_3 G\left(1;\bar{y}\right) 
           + 8 \left[ G\left(1,1,1;\bar{y}\right) + \zeta_2 G\left(1;\bar{y}\right) \right] \left[ G\left(1,\bar{x}\right) -2 \ln\left(\bar{x}\right) \right] 
   \right\} \eps^4
 + {\mathcal O}\left(\eps^5\right).
\eq
The function $G(z_1,\dots,z_r;\lambda)$ denotes a multiple polylogarithm.
These functions are defined as follows: If $z_1,\dots,z_r$ are all zero, we define $G(z_1,\dots,z_r;\lambda)$ by
\bq
 G(\underbrace{0,\dots,0}_{r-\mathrm{times}};\lambda) & = & \frac{1}{r!} \ln^r\left(\lambda\right).
\eq
This definition includes the trivial case $r=0$, where $G(;\lambda) = 1$.
If at least one variable $z$ is non-zero, we define recursively
\bq
 G\left(z_1,z_2\dots,z_r;\lambda\right)
 & = &
 \int\limits_0^\lambda \frac{d\lambda_1}{\lambda_1-z_1} G\left(z_2\dots,z_r;\lambda_1\right).
\eq
The variables $\bar{x}$ and $\bar{y}$ appearing in eq.~\eqref{eq_J37} are defined by
\bq
 \frac{\left(-s\right)}{m_1^2} \; = \; \frac{\bar{x}^2}{\left(1-\bar{x}\right)},
 & &
 \frac{\left(-t\right)}{m_1^2} \; = \; \frac{\bar{y}^2}{\left(1-\bar{y}\right)}.
\eq
Moreover, we substitute $m_2^2$ with 
\bq
 \frac{m_2^2}{m_1^2} & = &
 \frac{1-\bar{y}-\bar{y}^2}{\left(1-\bar{y}\right)} 
 + i \frac{\bar{y}}{\sqrt{1-\bar{y}}} \left( \tilde{z} + \frac{1}{\tilde{z}} \right),
\eq
and integrate in $\tilde{z}$ instead of $z$. The $\tilde{z}$-integration (or the $z$-integration) involves only the differential one-forms $(\omega_7-\omega_4)$ and $\omega_{12}$, with $\omega_{12}$ containing the root $r_3$. Substituting $m_2^2\to \tilde{z}$ rationalizes $r_3$ and allows us to express the iterated integrals in eq.~\eqref{eq_J37} as multiple polylogarithms. The new square root $i \sqrt{1-\bar{y}}$ is harmless, since $\bar{x}$ and $\bar{y}$ are constant parameters with respect to the $\tilde{z}$-integration (or $z$-integration).


\section{Conclusions}
\label{sect:conclusions}

In this article, we presented the results for all master integrals associated to the two-loop H-graph with unequal masses
in relativistic quantum field theory. We identified a total of $40$ master integrals for this family of Feynman integrals and constructed the $\epsilon$-factorized differential equation for them. The corresponding alphabet consists of $29$ differential one-forms in dlog-form, each of which is a dlog-form with either rational or algebraic arguments. The algebraic part consists of six simultaneously non-rationalizable square roots. Therefore, we presented the solution for all master integrals in terms of iterated integrals, where we used the known equal-mass case as the boundary condition. In addition, we present the H-graph with unit powers of the propagators up to weight four in terms of multiple polylogarithms in a in a rather compact way.

\section*{Acknowledgements}
This research was partly supported by the European Research Council (ERC)
under the European Union’s research and innovation programme grant agreements ERC
Starting Grant 949279 HighPHun.


\begin{appendix}


\section{Supplementary material}
\label{sect:supplement}

Attached to the arXiv version of this article is an electronic file 
\verb|supplementary_material.mpl|.
This file 
is in ASCII format with {\tt Maple} syntax, defining the quantities
\begin{center}
 \verb|A|, \; \verb|J|.
\end{center}
The matrix \verb|A| appears in the differential equation eq.~\eqref{diff_eq_J}
\bq
 d \vec{J} & = & \eps A \vec{J}.
\eq
The entries of the matrix $A$ are linear combinations of $\omega_1$, ..., $\omega_{29}$, defined in eqs.~\eqref{def_differential_one_forms}-\eqref{def_differential_one_forms_III}.
These differential forms are denoted by
\begin{center}
 \verb|omega_1|, ..., \verb|omega_29|.
\end{center}
The dimensional regularisation parameter $\eps$ is denoted by \verb|eps|.
The vector \verb|J| contains the results for the master integrals up to order $\eps^4$ in terms of iterated integrals.
An iterated integral $I_\gamma(\omega_1,\omega_2,\omega_3;z)$ for a path along the positive $z$-axis
from the starting point $0$ to the end point $z$ is denoted by
\begin{center}
 \verb|iter_int([omega_1,omega_2,omega_3],z)|.
\end{center}
The boundary values at $z=0$ are expressed in terms of multiple polylogarithms, and the notation of
ref.~\cite{Kreer:2021sdt} is used.


\section{Boundary data}
\label{sect:boundary}

The equal mass case $m_1=m_2=m$ provides convenient boundary values.
Up to weight four, we may express all master integrals for the equal-mass H-graph in terms of multiple polylogarithms.
The equal-mass case has $25$ master integrals. In the following, we denote the basis of master integrals for the equal-mass case 
given in ref.~\cite{Kreer:2021sdt}
by $J^{\mathrm{eq}}=(J^{\mathrm{eq}}_1,\dots,J^{\mathrm{eq}}_{29})$.
The basis of the unequal-mass case is denoted as $J=(J_1,\dots, J_{40})^T$.
The boundary values are:
\begin{align}
 \lim\limits_{m_1=m_2=m} J_{1} & = J_{1}^{\mathrm{eq}},
 &
 \lim\limits_{m_1=m_2=m} J_{2} & = J_{2}^{\mathrm{eq}},
 &
 \lim\limits_{m_1=m_2=m} J_{3} & = J_{2}^{\mathrm{eq}},
 &
 \lim\limits_{m_1=m_2=m} J_{4} & = J_{3}^{\mathrm{eq}},
 \nonumber \\
 \lim\limits_{m_1=m_2=m} J_{5} & = J_{4}^{\mathrm{eq}},
 &
 \lim\limits_{m_1=m_2=m} J_{6} & = J_{3}^{\mathrm{eq}},
 &
 \lim\limits_{m_1=m_2=m} J_{7} & = 2 J_{5}^{\mathrm{eq}},
 &
 \lim\limits_{m_1=m_2=m} J_{8} & = J_{6}^{\mathrm{eq}},
 \nonumber \\
 \lim\limits_{m_1=m_2=m} J_{9} & = J_{6}^{\mathrm{eq}},
 &
 \lim\limits_{m_1=m_2=m} J_{10} & = J_{7}^{\mathrm{eq}},
 &
 \lim\limits_{m_1=m_2=m} J_{11} & = J_{8}^{\mathrm{eq}},
 &
 \lim\limits_{m_1=m_2=m} J_{12} & = J_{8}^{\mathrm{eq}},
 \nonumber \\
 \lim\limits_{m_1=m_2=m} J_{13} & = J_{9}^{\mathrm{eq}},
 &
 \lim\limits_{m_1=m_2=m} J_{14} & = J_{10}^{\mathrm{eq}},
 &
 \lim\limits_{m_1=m_2=m} J_{15} & = J_{11}^{\mathrm{eq}},
 &
 \lim\limits_{m_1=m_2=m} J_{16} & = J_{12}^{\mathrm{eq}},
 \nonumber \\
 \lim\limits_{m_1=m_2=m} J_{17} & = J_{13}^{\mathrm{eq}},
 &
 \lim\limits_{m_1=m_2=m} J_{18} & = J_{10}^{\mathrm{eq}},
 &
 \lim\limits_{m_1=m_2=m} J_{19} & = J_{11}^{\mathrm{eq}},
 &
 \lim\limits_{m_1=m_2=m} J_{20} & = J_{12}^{\mathrm{eq}},
 \nonumber \\
 \lim\limits_{m_1=m_2=m} J_{21} & = J_{13}^{\mathrm{eq}},
 &
 \lim\limits_{m_1=m_2=m} J_{22} & = J_{9}^{\mathrm{eq}},
 &
 \lim\limits_{m_1=m_2=m} J_{23} & = J_{14}^{\mathrm{eq}},
 &
 \lim\limits_{m_1=m_2=m} J_{24} & = J_{14}^{\mathrm{eq}},
 \nonumber \\
 \lim\limits_{m_1=m_2=m} J_{25} & = J_{15}^{\mathrm{eq}},
 &
 \lim\limits_{m_1=m_2=m} J_{26} & = J_{16}^{\mathrm{eq}},
 &
 \lim\limits_{m_1=m_2=m} J_{27} & = J_{17}^{\mathrm{eq}},
 &
 \lim\limits_{m_1=m_2=m} J_{28} & = J_{18}^{\mathrm{eq}},
 \nonumber \\
 \lim\limits_{m_1=m_2=m} J_{29} & = J_{19}^{\mathrm{eq}},
 &
 \lim\limits_{m_1=m_2=m} J_{30} & = J_{20}^{\mathrm{eq}},
 &
 \lim\limits_{m_1=m_2=m} J_{31} & = J_{20}^{\mathrm{eq}},
 &
 \lim\limits_{m_1=m_2=m} J_{32} & = J_{21}^{\mathrm{eq}},
 \nonumber \\
 \lim\limits_{m_1=m_2=m} J_{33} & = J_{15}^{\mathrm{eq}},
 &
 \lim\limits_{m_1=m_2=m} J_{34} & = J_{16}^{\mathrm{eq}},
 &
 \lim\limits_{m_1=m_2=m} J_{35} & = J_{17}^{\mathrm{eq}},
 &
 \lim\limits_{m_1=m_2=m} J_{36} & = J_{22}^{\mathrm{eq}},
 \nonumber \\
 \lim\limits_{m_1=m_2=m} J_{37} & = J_{23}^{\mathrm{eq}},
 &
 \lim\limits_{m_1=m_2=m} J_{38} & = J_{24}^{\mathrm{eq}},
 &
 \lim\limits_{m_1=m_2=m} J_{39} & = J_{24}^{\mathrm{eq}},
 &
 \lim\limits_{m_1=m_2=m} J_{40} & = J_{25}^{\mathrm{eq}}.
\end{align}


\section{The integral $J_{28}$}
\label{sect:J_28}

Up to weight four the root $r_6$ enters only the master integral $J_{28}$.
\begin{figure}
\begin{center}
\includegraphics[scale=1.0]{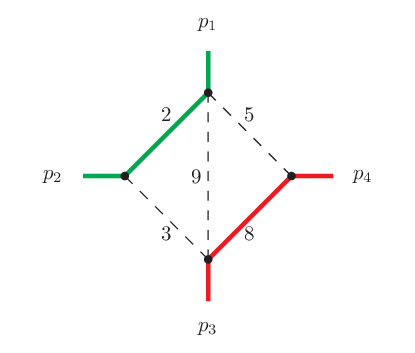}
\end{center}
\caption{
The topology for the master integral $J_{28}$.
This topology has five master integrals $J_{28}$-$J_{32}$.
Up to weight four the square root $r_6$ enters only $J_{28}$.
}
\label{fig_J28}
\end{figure}
The topology of this master integral is shown in fig.~\ref{fig_J28}.
The Feynman integral $J_{28}$ starts at ${\mathcal O}(\eps^4)$ and usually, we are only interested in terms up to weight four, 
hence we only need to compute $J_{28}^{(4)}$.
The integral $J_{28}$ is an example of a highly non-trivial Feynman integral, which can be computed alternatively to the method
of differential equations by direct integration in Feynman parameter space (for a review of this method, see ref.~\cite{Bourjaily:2021lnz}).
It is worth outlining this calculation, following the lines of ref.~\cite{Heller:2019gkq}.
We start from the Feynman parameter representation:
\bq
 J_{28}
 & = &
 4 \eps^4 r_6 I_{011010011}
 \nonumber \\
 & = &
 4 \eps^4 r_6 
 e^{2 \gamma_E \eps}
 \Gamma\left(1+2\eps\right)
 \int\limits_{\alpha_j \ge 0} d^5\alpha \delta\left(1-\alpha_8\right)
 {\mathcal U}^{-1+3\eps}
 {\mathcal F}^{-1-2\eps}.
\eq
The two graph polynomials are given by
\bq
 {\mathcal U}
 & = &
 \left(\alpha_2+\alpha_3\right) \left(\alpha_5+\alpha_8\right)
 + \left(\alpha_2+\alpha_3\right) \alpha_9
 + \left(\alpha_5+\alpha_8\right) \alpha_9,
 \nonumber \\
 {\mathcal F}
 & = &
 \alpha_3 \alpha_5 \alpha_9 \left(-s\right)
 + \alpha_2 \alpha_8 \alpha_9 \left(-t\right)
 + \alpha_2 \left[ \alpha_2 \left(\alpha_5+\alpha_8+\alpha_9\right) + \alpha_8 \alpha_9 \right] m_1^2
 \nonumber \\
 & &
 + \alpha_8 \left[ \alpha_8 \left(\alpha_2+\alpha_3+\alpha_9\right) + \alpha_2 \alpha_9 \right] m_2^2.
\eq
The integral starts at $\eps^4$ and we are only interested in $J_{28}^{(4)}$.
We have
\bq
 J_{28}^{(4)}
 & = &
 4 r_6 
 \int\limits_{0}^\infty d\alpha_2
 \int\limits_{0}^\infty d\alpha_5
 \int\limits_{0}^\infty d\alpha_9
 \int\limits_{0}^\infty d\alpha_3
 \int\limits_{0}^\infty d\alpha_8
 \frac{\delta\left(1-\alpha_8\right)}{{\mathcal U}{\mathcal F}}.
\eq
We perform the integrations right to left, starting with $\alpha_8$.
The integration over $\alpha_8$ and $\alpha_3$ are trivial, yielding
\bq
 I_1 & = & 
 \int\limits_{0}^\infty d\alpha_3
 \int\limits_{0}^\infty d\alpha_8
 \frac{\delta\left(1-\alpha_8\right)}{{\mathcal U}{\mathcal F}}
 \; = \;
 \frac{1}{{\mathcal P}_5}
 \ln\left( \frac{ {\mathcal P}_1 {\mathcal P}_2 }
                { {\mathcal P}_3 {\mathcal P}_4 }\right),
\eq
with
\bq
 {\mathcal P}_1
 & = & 
 1+\alpha_5+\alpha_9,
 \\
 {\mathcal P}_2
 & = &
 \alpha_2\left(\alpha_2\alpha_5 + \alpha_2 + \alpha_9 + \alpha_2 \alpha_9\right) m_1^2 
 + \left(\alpha_2+\alpha_9+\alpha_2\alpha_9\right) m_2^2 
 - \alpha_2\alpha_9 t,
 \nonumber \\
 {\mathcal P}_3 
 & = &
 m_2^2-\alpha_5\alpha_9 s,
 \nonumber \\
 {\mathcal P}_4 
 & = &
 \alpha_2 \alpha_9 + \left(\alpha_2+\alpha_9\right)\left(1+\alpha_5\right),
 \nonumber \\
 {\mathcal P}_5
 & = &
 \alpha_5 \alpha_9 \left[ \alpha_2 \alpha_9 + \left(\alpha_2+\alpha_9\right)\left(1+\alpha_5\right)\right] s
 - \alpha_2 \alpha_9 \left( 1+\alpha_5+\alpha_9 \right) t
 \nonumber \\
 & &
 + \alpha_2 \left( 1+\alpha_5+\alpha_9 \right) \left(\alpha_2\alpha_5 + \alpha_2 + \alpha_9 + \alpha_2 \alpha_9\right) m_1^2
 + \alpha_9 \left(\alpha_2\alpha_5 + \alpha_2 + \alpha_9 + \alpha_2 \alpha_9\right) m_2^2.
 \nonumber
\eq
The variable change
\bq
 \alpha_2 & = & \frac{x_2 \alpha_9}{1+\alpha_5+\alpha_9}
\eq
leads to
\bq
 \int\limits_{0}^\infty d\alpha_2
 \int\limits_{0}^\infty d\alpha_5
 \int\limits_{0}^\infty d\alpha_9
 \; 
 I_1
 & = &
 \int\limits_{0}^\infty dx_2
 \int\limits_{0}^\infty d\alpha_5
 \int\limits_{0}^\infty d\alpha_9
 \frac{1}{\alpha_9 {\mathcal P}_1 {\mathcal P}_7} \ln\left(\frac{{\mathcal P}_6}{{\mathcal P}_3 \left(1+x_2+a_5\right)}\right)
\eq
with
\bq
 {\mathcal P}_6
 & = &
 - x_2 \alpha_9 t + x_2 \alpha_9 \left(1+x_2\right) m_1^2 + \left(1 + x_2 + \alpha_5 + \alpha_9 + x_2 \alpha_9 \right) m_2^2,
 \nonumber \\
 {\mathcal P}_7 
 & = &
 \alpha_5 \left(1+x_2+\alpha_5\right) s
 - x_2 t
 + x_2 \left(1+x_2\right) m_1^2
 + \left(1+x_2\right) m_2^2.
\eq
We may now integrate out $\alpha_9$.
This gives
\bq
 I_2
 & = &
 \int\limits_{0}^\infty d\alpha_9
 \frac{1}{\alpha_9 {\mathcal P}_1 {\mathcal P}_7} \ln\left(\frac{{\mathcal P}_6}{{\mathcal P}_3 \left(1+x_2+a_5\right)}\right)
 \nonumber \\
 & = & 
 \frac{1}{2\left(1+\alpha_5\right) {\mathcal P}_7}
 \left\{
  \ln^2\left(\frac{\left(1+x_2+\alpha_5\right)m_2^2}{{\mathcal P}_8}\right)
  - \ln^2\left(\frac{m_2^2}{\alpha_5\left(-s\right)}\right)
  + 2 G\left(0,\frac{\alpha_5 s}{m_2^2+\alpha_5 s}; 1 \right)
 \right. \nonumber \\
 & & \left.
  - 2 G\left(0,\frac{{\mathcal P}_8}{{\mathcal P}_9};1\right)
  + 2 G\left(\frac{\alpha_5 s}{m_2^2+\alpha_5 s},-\frac{1}{\alpha_5}; 1 \right)
  - 2 G\left(\frac{{\mathcal P}_8}{{\mathcal P}_9},-\frac{1}{\alpha_5};1\right)
 \right\},
\eq
with
\bq
 {\mathcal P}_8
 & = &
 - x_2 t + x_2 \left(1+x_2\right) m_1^2 + \left(1+x_2\right) m_2^2,
 \nonumber \\
 {\mathcal P}_9
 & = &
 - x_2 t + x_2 \left(1+x_2\right) m_1^2 - \alpha_5 m_2^2.
\eq
The variable change
\bq
 m_2^2 & = & \frac{1}{4} \left( \frac{r_2^2}{\left(-s\right)} + s \right)
\eq
allows us to convert the multiple polylogarithms to a form, where $\alpha_5$ appears only as the upper integration limit.
The polynomial ${\mathcal P}_7$ is quadratic in $\alpha_5$.
Factorising ${\mathcal P}_7$ introduces the roots of ${\mathcal P}_7$:
\bq
\label{root_alpha_5}
 {\mathcal P}_7
 & = &
 s \left( \alpha_5 - \alpha_5^{(1)} \right) \left( \alpha_5 - \alpha_5^{(2)} \right),
 \\
 \alpha_5^{(1/2)}
 & = &
 - \frac{1}{2}\left(1+x_2\right)
 \pm \frac{1}{2s} \sqrt{-s\left(4m_1^2-s\right)x_2^2 -2s\left(2m_1^2+2m_2^2-2t-s\right)x_2 -s\left(4m_2^2-s\right)}.
 \nonumber
\eq
This allows us to integrate out $\alpha_5$.
This leads to
\bq
 J_{28}^{(4)}
 & = &
 4 r_6 
 \int\limits_{0}^\infty dx_2
 \frac{1}{2 s \left( \alpha_5^{(1)} - \alpha_5^{(2)} \right) \left(1+\alpha_5^{(1)}\right) \left(1+\alpha_5^{(2)}\right)}
 R,
\eq
with a remainder function $R$ of weight $3$.
The transformation
\bq
 x_2 
 & = &
 4
 \frac{y_2}{1-y_2}
 \frac{r_2^2}{\left[ \left(r_1+r_2\right)^2 + 4st - \left( \left(r_1-r_2\right)^2+4st\right) y_2\right]}
\eq
rationalises the root in eq.~\eqref{root_alpha_5}:
\bq
\lefteqn{
 \sqrt{-s\left(4m_1^2-s\right)x_2^2 -2s\left(2m_1^2+2m_2^2-2t-s\right)x_2 -s\left(4m_2^2-s\right)}
 = } & & 
 \nonumber \\
 & &
 \frac{r_2}{\left(1-y_2\right)}
 \frac{\left[ \left(r_1+r_2\right)^2 + 4st - \left( \left(r_1-r_2\right)^2+4st\right) y_2^2\right]}{\left[ \left(r_1+r_2\right)^2 + 4st - \left( \left(r_1-r_2\right)^2+4st\right) y_2\right]}.
\eq
The integration domain for $y_2$ is the interval $[0,1]$.
We convert $R$ to a form, where $y_2$ appears only as the upper integration limit.
This introduces two additional square roots
\bq
 r_7
 & = &
 \sqrt{\left(-s\right)\left(-t\right)},
 \nonumber \\
 r_8
 & = &
 \sqrt{4 m_1^2 m_2^2 - s t},
\eq
but -- more importantly -- allows us to express the integral $J_{28}^{(4)}$ in terms of multiple polylogarithms.
This calculation shows how the algorithm of \verb|HyperInt|~\cite{Panzer:2014caa} can be tweaked to accommodate
square roots.
 
\end{appendix}

{\footnotesize
\bibliography{/home/stefanw/notes/biblio}
\bibliographystyle{/home/stefanw/latex-style/h-physrev5}
}

\end{document}